\documentclass[12pt, reqno, oneside,amsart]{article}
\usepackage{geometry}
\usepackage{amsmath}

\input epsf

\usepackage[shortlabels]{enumitem}

\usepackage{cite}
\usepackage{graphicx}
\usepackage{amsfonts}
\usepackage{amssymb}

\usepackage{booktabs}

\usepackage{etoolbox}
\usepackage{titlesec}
\usepackage{titletoc}
\usepackage{titletoc}

 \usepackage[utf8]{inputenc}

\usepackage[titletoc]{appendix}

\usepackage{enumitem}

\usepackage[dvipsnames]{xcolor}

\usepackage{hyperref}
\hypersetup{
    colorlinks=true,
    anchorcolor=red,
    urlcolor=CadetBlue,
    citecolor=MidnightBlue,
    linkcolor=MidnightBlue}
    
\usepackage{empheq}

\usepackage{mathrsfs}  

\usepackage{dsfont}

\usepackage{braket}

\usepackage{wrapfig}

\usepackage{stmaryrd}

\usepackage{subfigure}

\numberwithin{equation}{section}

\titleformat{\section} 
[hang]                 
	{
	\filcenter
	\normalsize
	\large
	} 
{\thesection.}       
{1em}                  
{}

\titlespacing{\section}
{0pt}                  
{20pt}                  
{13pt}                  

\titleformat{\subsection} 
[hang]                 
{
\itshape
\filcenter
} 
{\thesubsection}       
{1em}                  
{}                     

\titlespacing{\subsection}
{0pt}                  
{20pt}                  
{15pt}                  

\titleformat{\subsubsection} 
[hang]
{
\filcenter
\normalsize
\itshape
} 
{\thesubsubsection}    
{1em}                
{}                     

\titlespacing{\subsubsection}
{0pt}                  
{25pt}		        
{15pt}                  

\def\n{\nu}
\def\a{\alpha}

\def\la{\lambda}

\def\ka{\varkappa}

\def\ga{\gamma}

\def\vare{\varepsilon}

\def\rm{\mathrm}

\def\scr{\mathscr}

\def\be{\begin{equation}}
\def\ee{\end{equation}}
\def\br{\begin{eqnarray}}

\def\er{\end{eqnarray}}
\def\bsub{\begin{subequations}}
\def\esub{\end{subequations}}

\def\IR{\mathrm{IR}}
\def\UV{\mathrm{UV}}

\usepackage{authblk}

\title{SUSY shields the scaling symmetry of conformal quantum mechanics}

\author[a]{\normalsize A.A. Lima\thanks{andrealves.fis@gmail.com}}
\author[a]{J.V.S. Scursulim\thanks{josevictor.s.scursulim@gmail.com}}
\author[a]{U. Camara da Silva\thanks{ulyssescamara@gmail.com}}
\author[a]{G.M. Sotkov\thanks{gsotkov@gmail.com}}

\affil[a]{\textit{\small Department of Physics, Federal University of Esp\'irito Santo}}

\begin{document}

\maketitle

\begin{abstract}

Renormalization of the inverse square potential usually breaks its classical conformal invariance. In a strongly attractive potential, the  scaling symmetry is broken to a discrete subgroup while, in a strongly repulsive potential, it is preserved at  quantum level. In the intermediate, weak-medium range of the coupling, an anomalous length scale appears due to a flow of the renormalization group away from a critical point. We show that potentials with couplings in the strongly-repulsive and in the weak-medium ranges can be related by a dynamical supersymmetry. Imposing SUSY invariance unifies these two ranges, and fixes the anomalous scale to zero, thus restoring the continuous scaling symmetry.


\end{abstract}

\tableofcontents

\section{Introduction}

The Hamiltonian of conformal quantum mechanics \cite{deAlfaro:1976vlx},
\be
H = \frac{1}{2m}\,  p^2 + V(x) \ ,
\quad
V(x) = \frac{\hbar^2}{2m} \frac{\alpha}{x^2} ,
	\label{IinvSqPot}
\ee
being singular, is infamously subtle \cite{Case:1950an,Frank:1971xx}. Yet, it appears in a cornucopia of physical problems, including the Efimov effect of nuclear 3-body scattering and its generalizations in condensed matter theory \cite{Efimov, moroz2015generalized};  mathematical physics \cite{Calogero:1970nt,Moser:1975qp,sutherland1971exact};  quantum field theory near the horizons of black holes
\cite{Govindarajan:2000ag,Birmingham:2001qa,Gupta:2001bg};
 fluctuations in gauge/gravity duality \cite{Gursoy:2007er,Kiritsis:2006ua,Kiritsis:2016kog,Lima:2019bsi}; the AdS$_2$/CFT$_1$ correspondence \cite{Chamon:2011xk}; and still other phenomena.
Such a breath of applications could be seen as a reflection of scaling invariance, which can appear as a (usually asymptotic) symmetry in various situations.

Scale invariance of (\ref{IinvSqPot}) is a consequence of the homogeneous transformation of $H$ under a rescaling
\be
t \to \rho^2 t , \qquad 
x \to \rho x , 
		\label{ConfrSymmt}
\ee
with $\rho > 0$. Since the momentum transforms as $p \to \rho^{-1} p$, 
the Hamiltonian has a definite dimension,  $H \to \rho^{-2} H$.
Thus (\ref{ConfrSymmt}) is a symmetry of the stationary Schr\"odinger equation
\be
\left[ -\frac{d^2}{dx^2}+\frac{\alpha}{x^2} \right] \psi(x) = \frac{2m}{\hbar^2} E \, \psi(x), \quad x > 0 , \label{Schr}
\ee
where $\psi(x) \to \tilde \psi(\rho x) = \rho^{-1/2} \psi( x)$, preserving the probability $dx|\psi(x)|^2$, and the energy changes as $E \to \rho^{-2} E$ \cite{Gitman}. 
But the symmetry can be broken at the level of the quantum states, as the regularization of the singularity of $V(x)$ may introduce ``anomalous'' length scales.

Breaking of scale invariance depends subtly on the value of the adimensional coupling $\a$. 
Na\"ively, there are two qualitatively distinct possibilities: either $\a > 0$ and the potential is repulsive, 
or $\a < 0$ and the potential is attractive. 
The latter is evidently problematic because the  singularity at $x = 0$ makes the question of wether the particle can ``fall to the center''  nontrivial \cite{Landau_3}.
Na\"ivet\'e is due precisely  to the singularity: the Hamiltonian (\ref{IinvSqPot}) is not self-adjoint, and physical results require a self-adjoint extension \cite{Gitman,Gitman:2009era,Derezinski:2016pey}. 
Constructing these extensions turns out to be completely equivalent to a renormalization procedure.
Strictly, the singularity  of $V(x)$ at $x = 0$ should be considered an effect of inadvertently extending the problem too much into the realm of some unknown short-distance physics.
Once regarding a singular potential such as (\ref{IinvSqPot})  as  an effective theory valid only at long distances, the singular vicinity of $x = 0$ requires  a renormalization procedure, to which observables at large $x$ should be insensitive \cite{Beane:2000wh,Bawin:2003dm,Camblong:2003mb,Braaten:2004pg,Kaplan:2009kr,Bouaziz:2014wxa,Bulycheva:2014twa,CamaradaSilva:2018leo}. 

The renormalized theory depends not simply on wether $\a$ is positive or negative, but on \emph{three} qualitatively different regimes:
\bsub\begin{flalign}
\text{Strongly Repulsive:}		&&	 \a &\in [\tfrac{3}{4} , \infty) &&	\label{StrgRepRang}
\\
\text{Weak Medium:}		&&	 \a &\in [- \tfrac{1}{4} , \tfrac{3}{4} ) &&	\label{WeakMedRang}
\\
\text{Strongly Attractive:}		&&	 \a &\in (- \infty , - \tfrac{1}{4} ) &&	\label{StrongAttrRang}
\end{flalign}\label{3Ranges}\esub
In the strongly repulsive range (\ref{StrgRepRang}), the renormalized solutions are scale-invariant, while in the strongly attractive range (\ref{StrongAttrRang}) scale invariance is broken into a discrete subgroup, and conformality is lost after a BKT-like phase transition happens at $\a = - \frac{1}{4}$ \cite{Kaplan:2009kr}.  
In the weak-medium range (\ref{WeakMedRang}), renormalization introduces the anomalous scale $L$, and the continuous family of self-adjoint extensions of (\ref{IinvSqPot}) corresponds to the renormalization group (RG) flow between two conformal fixed points where $L = 0$ and $L = \infty$. 
Therefore, in the weak-medium range, 
for finite $L$, conformality is \emph{also} lost  by  `dimensional transmutation' \cite{Camblong:2000qn,Coleman:1973jx,huang1992quarks}.
Nevertheless, since there is still the possibility of restoring continuous scaling symmetry by choosing one of the fixed points of the RG flow, we call the entire range of $\a \in [- \frac{1}{4} , \infty)$ the `continuous-scaling phase', in contrast with the `discrete-scaling phase' of $\a < - \frac{1}{4}$.

The objective of the present paper is to show that the continuous-scaling phase has a somewhat disguised symmetry that unifies the strongly-repulsive and the weak-medium ranges:  a supersymmetry (SUSY) of the inverse square potential. 
This is \emph{not} an extension of the 1D conformal algebra, such as the ones which have been considered e.g. in the context of holography of  black holes \cite{Claus:1998ts}. Rather, it is a dynamical symmetry of the energy spectrum due to the factorization  \cite{Cooper:1994eh} of  the Hamiltonian (\ref{IinvSqPot}) into  two  different products  
$H_+ = \frac{\hbar^2}{2m} Q^\dagger Q  \;\; \text {and} \;\;\;  H_- = \frac{\hbar^2}{2m} Q Q^\dagger$, where  
\[
 Q = \frac{d}{dx} + \frac{\sqrt{2m}}{\hbar} W(x) \qquad \text{and} \qquad \frac{\sqrt{2m}}{\hbar}W(x) = -\frac{ \nu + \frac{1}{2} }{x} ,
 \]
producing a pair of  inverse square potentials $V_+(x)$ and  $V_-(x)$ with different couplings $\a_\pm$, determined by $\a = \nu^2 - \frac{1}{4}$, with $\nu_- = \nu_+ +1$.
Our main observation is that consistency with this supersymmetry forces the anomalous scale in the weak-medium range to \emph{vanish}, restoring conformal symmetry over the whole continuous-scaling phase. 
In the discrete-scaling phase, the SUSY construction leads to inverse-square potentials with complex couplings so, in this sense, it ceases to be a symmetry of (\ref{IinvSqPot}).

In Sect.\ref{SectRenor}, we review the renormalization procedure for the inverse square potential, and how the anomalous scale appears in the weak-medium coupling. In Sect.\ref{sec:SUSY} we review the basic aspects of SUSY quantum mechanics, and show our main result. In Sect.\ref{sec:Apli} we present a collection of examples and show how our construction can be generalized to supersymmetric potentials which are only asymptotically like (\ref{IinvSqPot}). We conclude with a brief discussion.

\section{Conformal symmetry and renormalization of the inverse square potential}	\label{SectRenor}

For $E > 0$ the general solution of (\ref{IinvSqPot}) is
\be
\begin{split}
\psi_{\n ,k} (x) = A_{\nu, k} \ \sqrt{x} J_{\nu}(kx) + B_{\nu,  k} \ \sqrt{x} N_{\nu}(kx), 
\\
\nu \equiv \sqrt{ \a + 1 / 4 } ;
\quad
k \equiv \sqrt{2m E / \hbar^2} .
\end{split}	 \label{sol}
\ee
with $A_{\nu, k}$ and $B_{\nu, k}$ integration constants. 
 The three ranges of $\a$ translate to $\n$ as in Table \ref{Tabalphtonu}.
We most often use the index $\n$ instead of $\a$. 
For now, consider $\n \geq 0$ and leave the discussion of the  strongly attractive range, characterized by an imaginary $\n = i \tilde \nu$, for later in this section.

Physical wave functions must be \emph{normalizable} in the vicinity of the singular point,
\be
\lim_{x_0 \to 0} \int^{x_0} \! \! dx \, | \psi_{\n;k}(x)|^2 < \infty .	\label{norm}
\ee
The first solution in (\ref{sol}) is always square-integrable at $x = 0$, since $J_\nu(kx) \sim (kx)^\n$. The second solution goes as  $N_{\nu}(kx)\sim (kx)^{-\nu}$ for $kx\ll1$, so its norm diverges for $\n \geq 1$ and, therefore, in the strongly-repulsive range normalizability fixes
\be
B_{\nu , k} =0 \quad \text{if} \quad \nu \geq 1 \quad \text{(Strongly Attractive)}.	\label{StrRepB0}
\ee
Hence the wave-function is determined uniquely ($A_\n$ just fixes the norm). 
On the other hand, in the weak-medium range, $0 \leq \n < 1$, \emph{both} solutions in (\ref{sol}) are normalizable, so both constants $A_\n$ and $B_\n$ are arbitrary: the wave-function is not uniquely fixed.

\begin{table}
\begin{center}
\begin{tabular}{c c c}
		Str. Attractive & Weak Medium & Str. Repulsive
\\
\hline
\\
		$-\infty < \a < - \tfrac{1}{4}$ & $- \tfrac{1}{4} \leq \a < \tfrac{3}{4}$ & $\tfrac{3}{4} \leq \a < \infty$   
\\
		$-\infty < \nu^2 < 0$ & $0 \leq \nu^2 < 1$ & $1 \leq \n^2 < \infty$    
\end{tabular}
\end{center}
\caption{Qualiratively different ranges of the coupling: $\alpha$ versus $\nu$.}
\label{Tabalphtonu}
\end{table}

\subsection{Renormalization and the anomalous scale}

In any case, the singularity of $V(x)$ at the origin should be seen as the effect of using an effective theory outside its range of validity. 
Physical consistency can be obtained with a renormalization procedure: first, we define a regularized potential which is well-behaved at the origin; then we impose that physics at large distances should be insensitive to this regularization.
Essentially, these steps have all been presented elsewhere, cf. e.g. \cite{Beane:2000wh,Bouaziz:2014wxa,Bulycheva:2014twa,CamaradaSilva:2018leo}, but we outline them now for completeness and for fixing our notation. 
First define the regularized potential 
\be
\frac{2m}{\hbar^2}V_R(x) = \begin{cases}
					 \alpha / x^2, & x > R  
					 \\
					- \lambda / R^2, & x < R
					\end{cases}
\label{V_R}
\ee
The short-distance cutoff scale $R$ is much smaller than the only length scale of the system,  i.e. $k R \ll 1$, and 
we impose the Dirichlet condition $\psi(0) = 0$. A Neumann boundary condition would give equivalent results. Also, the use of a square well near the origin is just a convenient choice: other regularizations (e.g. a Dirac delta function) give equivalent results, as it should be \cite{Bouaziz:2014wxa}.

The regularization parameters $\la$ and $R$ must be related to each other in such a way that the long-distance properties of the system (including the coupling $\a$), are insensitive to a sliding of the cutoff.
The relation $\la(R)$ must be a property of the theory, hence it must be the same at any particular energy. We take as a reference the ``ground state'' (see, however, \S\ref{SectSUSYofISP}) solution with $E = 0$, which in the $x > R$ region is  simply
\be
\psi_{\nu,0} (x) = A_{\n,0}\,  x^{\frac{1}{2} +\nu} + B_{\n,0} \, x^{\frac{1}{2} - \nu} .
\label{Eeq0}
\ee
 It is clear that a length scale $L$, defined by 
\be
 L^\nu \equiv \vare B_{\n,0} / A_{\n,0}  , \quad \vare \equiv \rm{Sign} \big[ B_{\n,0} / A_{\n,0}  \big] = \pm 1 ,	\label{AnomSclL}
\ee
appears intrinsically into the solutions if both $A_{\n,0} , B_{\n,0} \neq 0$. 
In the regularized region the solution is 
$C_R \sin(\sqrt{\lambda}x/R)$, and imposing continuity of the logarithmic derivative $\psi'(x) / \psi(x)$ across the divide $x = R$ results in 
\be
\ga (R) \equiv \sqrt{| \la | } \cot \sqrt{| \la |} - \tfrac{1}{2}
		= \n \left[ \frac{1-\vare \left( L / R \right)^{2\n} }{ 1 + \vare \left(L / R \right)^{2\n} } \right]  \label{gamma} 
\ee
We do the same for a solution with $E > 0$, given by (\ref{sol}) for $x > R$ and by $C_\n \sin \ka x$ for $x < R$, where $\ka \equiv \sqrt{\la + (kR)^2} / R$. Using the leading asymptotic forms of the Bessel functions for arguments $kR \ll 1$, we find \cite{Bouaziz:2014wxa}
\be
\gamma(R) = \n \left[ \frac{1 + \frac{B_{\nu}}{A_{\nu}}\frac{\nu [\Gamma(\nu) ]^2}{\pi}\left(\frac{kR}{2}\right)^{-2\nu}}{1-\frac{B_{\nu}}{A_{\nu}}\frac{\nu [ \Gamma(\nu) ]^2}{\pi}\left(\frac{kR}{2}\right)^{-2\nu}} \right] .
	\label{gammaEpos}
\ee
Now we can combine Eqs.(\ref{gammaEpos}) and (\ref{gamma}) to obtain an explicit relation between the constants in terms of the anomalous scale $L$,
\be
\frac{ B_{\n,k}}{ A_{\n,k}} = - \frac{ \vare \pi}{\nu \left[ \Gamma(\nu) \right]^2 } \left(\frac{kL}{2}\right)^{2\n} .
			\label{BAneq0}
\ee
Remarkably, Eq.(\ref{BAneq0}) does not depend on $R$, which can be taken to zero, so we end up with a family of renormalized wave functions given by (\ref{sol}), parameterized by $L$ according to (\ref{BAneq0}).
Thus renormalization introduces a quantum anomalous scale, and conformal symmetry is spontaneously broken --- a `dimensional transmutation' \cite{Camblong:2000qn,Coleman:1973jx,huang1992quarks} happens.

If $\n = 0$ the regularized zero-energy wave function is
\be
\psi_{0,0}^{(R)}(x) = \begin{cases}
			\sqrt x \, \left[ 1+c \log(x/L_0) \right],  & x>R
			\\ 
			C_{0} \sin ( \sqrt{\la}x / R ), & x < R
			\end{cases}
	\label{psi_R_0}			
\ee
The anomalous scale, which we denote by $L_0$, here appears together with a dimensionless constant $c$. 
Imposing continuity of $\psi'/\psi$ at $x = R$ leads to a qualitatively different coupling,
\be
\gamma(R)  = \frac{c}{1+c \log(R/L_0)} \qquad (\n = 0)  \label{gamma_0}
\ee
and combining (\ref{gamma_0}) with (\ref{gamma}) we find the relation between the integration constants, to be compared with (\ref{BAneq0}), 
\be
\frac{A_{0,k}}{B_{0,k}} = - \frac{2}{\pi} \log \left( e^{-\frac{1}{c} + \scr C_E} k L_0 \right), 
	\label{ABok}
\ee
where $\scr C_E\approx 0.577$ is Euler's constant.

\subsection{RG flows}

The running of the regularization coupling $\ga(R)$  
can be seen as a renormalization group flow for the interaction \cite{Beane:2000wh,Bawin:2003dm,Braaten:2004pg,Kaplan:2009kr,Bouaziz:2014wxa,Bulycheva:2014twa,CamaradaSilva:2018leo}.
 In fact, $\ga(R)$ given in (\ref{gamma}) is the solution of the RG equation defined by the beta-function
\be
\beta_\ga \equiv \frac{d\ga}{d\log(R/L)} = -(\ga^2-\n^2).	\label{beta}
\ee
Its zeros define the ``ultraviolet'' and ``infrared'' fixed points
\be
\{ \ga_\UV = \n \ ; \  L = 0 \} 
\quad \text{and} \quad
\{\ga_\IR = - \nu \ ; \  L = \infty \} 	\label{UVIRfixpts}
\ee
where the coupling is independent of the cutoff and there is no anomalous scale.
In the strongly attractive regime (\ref{StrongAttrRang}) there are no zeros of $\beta_\ga$, as $\n = i \tilde \nu$ becomes imaginary; then the RG becomes cyclic \cite{Glazek:2002hq}.
This phase has a discrete energy spectrum $E_n \sim e^{-2\pi n / \bar \n}$, which is unbounded below and has an accumulation point near $E = 0$ \cite{Gitman}. Conformal symmetry is lost, but a discrete subgroup is preserved: (\ref{ConfrSymmt}) remains valid only for a  discrete set of scaling parameters $\rho = e^{\frac{\pi}{\bar\nu}n}$. 
We give some details of this process in the Appendix \ref{AppA}.
Now, we concentrate on the case of real $\nu$.
Thinking of $\n$ as an external parameter (corresponding to, say, a temperature)  we can consider what happens when it sweeps the interval $\nu \geq 0$, spanning the weak-medium and the strongly-repulsive regimes.

In the strongly repulsive regime ($\nu \geq 1$), the condition (\ref{StrRepB0}) fixes the scale $L = 0$, restricting the physical theory to the UV fixed point in (\ref{UVIRfixpts}).
There is no RG flow and scaling invariance is unbroken.

 Lowering $\nu$, we enter the weak-medium range ($0 \leq \nu < 1$). The situation complicates considerably.
 After imposing (\ref{BAneq0}), every theory with finite $L$ is equally physical, since $L$ is not required to vanish by normalizability.
 Thus the theory can leave the UV fixed point, with two possible fates of the RG flow: it can go to the IR fixed point (\ref{UVIRfixpts}), or it could develop a massive limit. 
 This latter case is subtle and less studied%
 \footnote{%
 See, however, the preprint version of \cite{CamaradaSilva:2018leo}, by one of the present authors.}
 so let us make a brief description.
The massive limit appears if the sign $\vare = -1$ in (\ref{AnomSclL}). Then from Eq.(\ref{gamma}) we see that the cutoff $R$ is restricted to the range $R \in ( 0 , L)$,%
\footnote{%
Technically we could also have, separately, the situation where $L < R$, but this is inconsistent with the condition that $R$ is smaller than any scale of the theory; in particular, we cannot take the limit $R \to 0$.}
since the function $\gamma(R)$ diverges at the finite scale $R = L$, the hallmark of a ``massive flow''. 
This flow is associated with a bound state:
if $E < 0$, the solution of (\ref{Schr}) which is square-integrable at $x = \infty$ is a modified Bessel function,
\[
\psi_{\n , \kappa}(x) = C \sqrt{\kappa x} K_{\nu}(\kappa x) , \quad  \kappa \equiv \sqrt{-2mE/\hbar^2}
.
\]
Continuity with the regularized region gives (using $\kappa R \ll 1$) 
\be
\kappa = 2\left[ - \tfrac{1}{\pi} \vare \nu \left[ \Gamma(\nu) \right]^2 \sin\pi\nu  \right]^{\frac{1}{2\nu}} L^{-1}\ , \quad  0< \n <1 .	\label{lig_1} 
\ee
If $\vare = -1$, there is a bound state with energy $E \sim - (\hbar^2 / m) L^{-2}$. 
It is not surprising to find a bound state because the regularized potential contains a well near the origin.%
\footnote{%
Alternatively, the bound state can be seen to appear because for $R \to 0$ the regularized potential has a delta function at the origin, which is known to support a bound state depending on the coupling.}
 But it is surprising that the state actually does not depend on $R$, so it persists even in the limit $R \to 0$. 
Actually, it can be paradoxical because it is also independent on the value of $\a$, so the bound state exists even if the potential is repulsive ($\a > 0$), or we have a free particle moving on the half-line ($\a = 0$).

As $\n \to 0^+$, so that $\a \to (-1/4)^+$, the two zeros of $\beta_\ga$ merge into a single fixed point
\be
\ga_\UV =  \ga_\IR \equiv \ga_{\rm{BKT}} = 0  .	\label{BKTga}
\ee
In \cite{Kaplan:2009kr}, it was shown that this is equivalent to a BKT-like phase transition.
Now, the solution of (\ref{beta}) is (\ref{gamma_0}).
If $c \neq 0$, we are outside the BKT fixed point (\ref{BKTga}), and there is again a massive RG flow. We can only take the limit $R \to 0$ if $c < 0$, and we find that this flow is associated with a bound state with energy 
\be
\kappa = 2  e^{\frac{1}{c} - \scr C_E} / L_0 \ , \quad \n = 0.
	\label{lig_2}
\ee
Again, the bound state is independent both of $R$, so it persists in the limit $R \to 0$.

\section{Supersymmetry in the continuous-scaling phase}\label{sec:SUSY}

The discussion in the previous section can be summarized as follows:
\begin{description}[style=unboxed,leftmargin=0cm]

\item[In the strongly attractive range ($\n = i \tilde \nu$),]
only a discrete subgroup of  scaling symmetry remains.

\item[In the strongly repulsive range ($\n \geq 1$),]
continuous scaling symmetry exists. Moreover, normalizability of the states imposes $B_{\n,k} = 0$, thus eliminating the anomalous scaling $L$ that appears in the renormalization process. 

\item[In the weak-medium range ($0 \leq \n < 1$),]
continuous scaling symmetry exists on the two fixed points (\ref{UVIRfixpts}) of the RG flow. 
But the theory is free to flow from the UV (or BKT) fixed point where $L = 0$, and develop a finite anomalous scale. The flow might go towards the IR fixed point where $L = \infty$ and scaling symmetry is restored; or, if $\vare = -1$, the flow might develop a massive limit associated with a bound state.
In any case, scaling symmetry is broken outside the fixed points. Worse, in the massive flows the bound state is definitely paradoxical if the potential is repulsive (or if $\a = 0$).

\end{description}

Here we come to the point of this paper, which is to show that the entire ``continuous-scaling phase'', consisting of the weak-medium plus the strongly-repulsive ranges, is unified by a  supersymmetry of the Hamiltonian (\ref{IinvSqPot}), which is destroyed by theories that flow away from the UV fixed point.%
\footnote{Or the self-adjoint extensions with $L > 0$.}
SUSY fixes $L = 0$ for every $\n \geq 0$, thus restoring scaling symmetry.
In particular, the energies of the bound states (\ref{lig_1}) and (\ref{lig_2}) become infinite when $L = 0$, and they are excised from the spectrum.

\subsection{SUSY quantum mechanics}

A non-relativistic quantum system is said to be `supersymmetric' if its Hamiltonian $H_+$ can be factored as 
\cite{Cooper:1994eh}
\be
H_+ = \frac{\hbar^2}{2m} Q^\dagger Q	\label{SUSYH1}
\ee
where the operator $Q$ and its conjugate $Q^\dagger$ are given by
\be
Q = \frac{d}{dx} + \frac{\sqrt{2m}}{\hbar} W(x), \quad Q^\dagger = - \frac{d}{dx} + \frac{\sqrt{2m}}{\hbar}W(x)
\ee
for a function $W(x)$, called the `superpotential', which determines the Schr\"odinger potential as
\be
\frac{2m}{\hbar^2} V_+ (x) = \Big[ \frac{\sqrt{2m}}{\hbar} W(x) \Big]^2 - \frac{\sqrt{2m}}{\hbar} \frac{d}{dx} W(x) .	\label{SUSYV1}
\ee
Given such a Hamiltonian $H_+$, there is a `partner' Hamiltonian $H_-$ with the inverse factorization, i.e.
\be
H_- = \frac{\hbar^2}{2m} Q Q^\dagger ,
\ee
whose corresponding potential has a flipped sign:
\be
\frac{2m}{\hbar^2} V_-(x) = \Big[ \frac{\sqrt{2m}}{\hbar} W(x) \Big]^2 + \frac{\sqrt{2m}}{\hbar} \frac{d}{dx} W(x) .
	\label{V2SUSYpar}
\ee
Supersymmetry stems from the fact that the matrix operators
\be
\scr H = \begin{bmatrix}
		H_+ & 0 
		\\
		0 & H_-
		\end{bmatrix} ,
\quad		
\scr Q = \begin{bmatrix}
		0 & 0 
		\\
		Q & 0
		\end{bmatrix} ,
\quad		
\scr Q^\dagger = \begin{bmatrix}
		0 & Q^\dagger 
		\\
		0 & 0
		\end{bmatrix} ,
\ee
form the closed superalgebra $\mathfrak{sl}(1|1)$, 
\be
 \{ \scr Q, \scr Q^\dagger\} = \scr H, 
\quad
[\scr H, \scr Q] = [\scr H, \scr Q^{\dagger}] = 0, 
\quad
 \{ \scr Q, \scr Q\} = \{\scr Q^{\dagger}, \scr Q^\dagger\} = 0. 
\ee 
These operators act on the space of `superstates' generated by
\be
\ket{\psi^{+}_k } = \begin{bmatrix}
				\psi^{+}_k \\ 0
			\end{bmatrix} ,	
\quad
\ket{\psi^{-}_k } = \begin{bmatrix}
				0 \\ \psi^{-}_k
			\end{bmatrix} ,
\ee	
where $\psi_k^{\pm}(x)$ are eigenstates of $H_\pm$,
\be
\frac{2m}{\hbar} H_\pm \psi_k^{\pm} = k^2 \psi_k^{\pm} , \quad A = 1,2.
\ee
The $\psi^{+}_k$ sector is said to be `bosonic', and the $\psi^{-}_k$ to be `fermionic', and we follow this nomenclature.
The factorization (\ref{SUSYH1}) relates the partner's spectra $\{E^{\pm} \}$ and their eigenvectors $\psi_k^{\pm}$. 
The vacum state, defined as
$\ket{0} = \ket{\psi^{+}_0} + \ket{\psi^{-}_0}$	, is annihilated by the charge $\scr Q$, viz.
$\scr Q \ket{0} = 0$.
 When the vacuum energy is zero, this implies that $\psi^{+}_0$ is a solution of $Q  \psi^{+}_0 = 0$,
\be
\psi^{+}_0(x) \sim \exp \big[ - \tfrac{\sqrt{2m}}{\hbar} \smallint dx \, W(x) \big] , \quad H_+ \psi_0^{+} = 0 .
\ee
If this function is square-integrable, then the vacuum lies completely in the bosonic sector, 
$\ket{0} = \ket{\psi^{+}_0}$, and the discrete energy spectra are related as $E^{-}_n = E^{+}_{n+1}$, with $n = 0,1,2,3, \cdots$. If, however, $\psi^{+}_0$ is \emph{not} square-integrable, then SUSY is said to be `spontaneously broken' and the energy spectra are completely degenerated, with the partner wave functions related by 
\be
\psi^{-}_{k}(x)=\frac{1}{k}Q\psi^{+}_{k}(x), \quad \psi^{+}_{k}(x)=\frac{1}{k}Q^{\dagger}\psi^{-}_{k}(x).\label{partners}
\ee
This will be our case of  interest. Note that the construction of SUSY partners is completely algebraic, and insensitive to wether one of the Hamiltonians should, eventually, not be self-adjoint.

\subsection{SUSY of the inverse square potential} 	\label{SectSUSYofISP}

The gist of this paper is that quantum mechanics with the inverse square potential is a SUSY quantum mechanics, with the superpotential
\be
\frac{\sqrt{2m}}{\hbar}W_\nu(x) = -\frac{ \nu + \frac{1}{2} }{x} ,
	\label{W}
\ee
associated with the operators $Q_\n$ and $Q_\n^\dagger$.
Indeed, the SUSY partners given by (\ref{SUSYV1})  and (\ref{V2SUSYpar}) are 
\be
\frac{2m}{\hbar^2} V_+(x) =  \frac{\nu^2 - \frac{1}{4}}{x^2} \ ,
\quad
\frac{2m}{\hbar^2} V_-(x) =  \frac{(\nu +1)^2 - \frac{1}{4}}{x^2} .	\label{H_SUSY_2}
\ee
Both are inverse square potentials, hence we say that (\ref{IinvSqPot}) is `shape invariant' under SUSY, whose effect is to change the coupling as
\be
\nu \mapsto \tilde \nu = \nu + 1 .		\label{tildenu}
\ee
Thus the partner of the BKT phase transition potential, which has $\a_+ = - \frac{1}{4}$, is the potential with $\a_- = \frac{3}{4}$, which is the lowest value of the coupling in the strongly repulsive range (see Table \ref{Tabalphtonu}).
In general, \emph{the entire weak-medium interval $\a \in [ - \frac{1}{4} , \frac{3}{4} )$ has been mapped by SUSY to the strongly repulsive interval $\a \in [ \frac{3}{4} , \frac{15}{4} )$.}

Looking at the eigenstates, first we see that the vacuum $\psi^{+}_{\n, 0}$, is such that 
\be
Q_\n \psi^{+}_{\n, 0} = 0, 
\quad
\text{hence}
\quad 
\psi^{+}_{\n, 0} = A_{\n,0} \, x^{\frac{1}{2} +\nu} .
		\label{SUSYvacISP}
\ee 
This function is not square-integrable on $[0, \infty)$. Therefore SUSY is spontaneously broken, the partner spectra are completely degenerated, and every eigenvector of $H_+$ is related to an eigenvector of $H_-$.
From Eq.(\ref{partners}) we can find the partners to the wave functions $\psi^{+}_{\n, k}(x)$ given in (\ref{sol}).
Making use of a recurrence formula for the Bessel functions%
\footnote{If $\scr C_{\nu}(z)$ is a solution of the Bessel equation with index $\nu$, then $\scr C_{\nu\pm1} = \mp \scr C_{\nu}'(z)+(\nu / z) \scr C_{\nu}(z)$ is a solution of the Bessel equation with index $\nu\pm1$; see \cite{NIST:DLMF} \S10.6(i).}
we find
\begin{align*}
\begin{split}
\psi^{-}_{\n, k} (x) &= \frac{1}{k}\left(\frac{d}{dx}-\frac{\nu+\frac{1}{2}}{x} \right) \psi^{+}_{\nu, k}(x)
	=\sqrt{x}\left[- A_{\nu ,k} J_{\nu+1}(kx) - B_{\nu,k} N_{\nu+1}(kx)\right]
\\
&\equiv\psi_{\nu+1,k}^{+} (x) . 
\end{split}		
\end{align*}
Indeed, the map (\ref{tildenu}) appears. Most importantly,  the integration constants are related by
\be
A_{\nu+1 , k} = - A_{\n, k} , \qquad B_{\nu+1 , k} = - B_{\nu, k}  .	\label{duais}
\ee
We emphasize that the fact that SUSY is spontaneously broken is crucial; if this were not the case, the relation (\ref{duais}) would not hold.

Eq.(\ref{duais}) has a remarkable consequence.
Suppose we start with a theory in the weak-medium range, with
\be
- \tfrac{1}{4} \leq \a_+ = \nu^2 - \tfrac{1}{4} < \tfrac{3}{4} .
\ee
We renormalize the theory, and the constants $A_{\n, k}$ and $B_{\n, k}$ are related by the anomalous scale $L$ as in Eq.(\ref{BAneq0}).
The (fermionic) partner model has the coupling
\be
\a_- = (\n + 1)^2 - \tfrac{1}{4} > \tfrac{3}{4},
\ee
which is in the strongly repulsive range, where normalizability fixes $B_{\n+1 , k} = 0$, cf. Eq.(\ref{StrRepB0}).
But then Eq.(\ref{duais}) forces us to make $B_{\n, k} = 0$, hence $L = 0$, thus \emph{selecting the UV point for the model in the medium-weak range}. 


 Hamiltonians in the weak-medium range have one single bound state with energy
$E \sim - 1/L^2$ given by (\ref{lig_1}) or (\ref{lig_2}). By fixing $L = 0$, thus $E = - \infty$,
these states are excised from the spectrum. This solves the paradox of bound states in repulsive or free potentials with $\a \in [0, \frac{3}{4})$. This is also consistent with the fact that there could be no bound states even in the attractive weak-medium range, $\a \in [-\frac{1}{4}, 0)$, because SUSY is a symmetry between the spectra of these models and the spectra of strongly-repulsive models with $\a \in [\frac{3}{4}, 2)$.

In the strongly-attractive range,  $\nu = i \tilde \nu$ becomes an imaginary number and the superpotential (\ref{W}) also becomes imaginary.
One could argue that the important thing would be for $V_+$ to be real, but
\begin{equation*}
\frac{2m}{\hbar^2} V_+(x) =  - \frac{(\tilde \nu^2 + \frac{1}{4})}{x^2} \ ,
\qquad
\frac{2m}{\hbar^2} V_-(x) =  \frac{\frac{3}{4} - \tilde \nu^2 + 2 i \tilde \nu}{x^2} 
\end{equation*}
so the partner potential $V_-(x)$ is complex and we have a ``non-Hermitian quantum mechanics'' \cite{Bender:2007nj,moiseyev2011non}. 
In this sense, SUSY is a property of the inverse square potential only in the range where scaling symmetry has not been discretely broken.

Since the inverse square potential is shape invariant,  we can repeat the procedure but now taking $V_-(x)$ as a \emph{bosonic} model, whose fermionic partner will have the Bessel index $\nu +2$. Going on like this generates a chain of models with indices $\nu + n$, all having the same spectrum and constants related as
\be
\begin{split}
B_{\nu+n , \, k} &= 0  
\\
A_{\nu+n+1 , \, k}  &= -  A_{\nu+n , \, k} = \cdots = (-1)^n  A_{\n ,k}  \ ; \qquad n \in \mathbb N .
\end{split}
\ee
Every model in this chain but the first one lies in the regime of strongly repulsive couplings.

For the purpose of illustration, let us consider explicitly the case $\n = 1$, $\a = \frac{3}{4}$. This is the smaller value of $\a$ and $\n$ inside the strongly-repulsive regime. The general renormalized energy eigenstate is (\ref{sol}), with the integration constants related by the anomalous scale according to (\ref{BAneq0}), that is
\be
\psi_{1,k}(x) = A_{1,k} \sqrt{ x} \left[  J_1(kx) - \vare \pi \left( \tfrac{1}{2} kL \right)^2   N_1(kx) \right] .
	\label{psi1k}
\ee
The function $\sqrt{k x} N_1(kx)$ is not square-integrable at $x = 0$ and so it must be absent, hence we must set $L = 0$ to make $\psi_{1,k}$ a good wave function. But let us carry on with $L \neq 0$ for a while.
Starting from $\psi_{1,k}$, we can obtain two different SUSY partners, depending on wether we take it to be the in the ``bosonic'' or in the ``fermionic'' sector. In the latter case, we have $\psi_{1,k} = \psi^{-}_{0,k}$, and we can recover the partner function $\psi^{+}_{0,k}$ by applying $Q_0^\dagger$ according to (\ref{partners}). The result is
\begin{align}
\begin{split}
\psi^{+}_{0,k} &= \frac{1}{k} Q_0^\dagger \psi^{-}_{0,k}
\\
	&= \frac{1}{k}  \left[ - \frac{d}{dx} - \frac{0 + 1/2}{x} \right] \psi_{1,k}(x) 
\\
	&= A_{1,k} \sqrt{x} \left[ - J_0(kx) + \vare \pi \left( \tfrac{1}{2} k L \right)^2 N_0(kx) \right]	
\end{split}
	\label{psi0kplu}
\end{align}
This is the general solution of the potential with $\nu = 0$, and $\a = - \tfrac{1}{4}$, the threshold of the strongly-attractive range. The function $\sqrt{x} N_0(kx)$ vanishes at $x = 0$, and looking only at (\ref{psi0kplu}) there is no reason to set $L = 0$. But now recall that if we do not set $L = 0$, then $\psi^{-}_{0,k}$ is not well-defined. Keeping track of $L$ before setting it to zero also reveals a subtlety of the $\n = 0$ solution.
Comparing the relative constant in the last line of (\ref{psi0kplu}) with formula (\ref{ABok}), one finds a non-trivial relation between the anomalous scales,
\be
- \tfrac{1}{\pi} \vare  \left( \tfrac{1}{2} k L \right)^{-2} 
		= - \tfrac{2}{\pi} \log \left( e^{-\frac{1}{c} + \scr C_E} k L_0 \right) 
		= - \tfrac{1}{\pi} \left[ -\tfrac{2}{c} + 2\scr C_E +\log( k^2 L_0^2) \right] .
\ee
Matching powers, we see that fixing $L = 0$ corresponds to fixing the dimensionless constant $c = 0$ in Eqs.(\ref{psi_R_0})-(\ref{ABok}),  not $L_0 = 0$. Of course, if $c = 0$ the coupling (\ref{gamma_0}) is fixed to lie on the BKT fixed point (\ref{BKTga}), and the length scale $L_0$ disappears.
Recall that the only way for the theory to flow away from the BKT fixed point where $c = 0$ is in the direction $c < 0$.

Next, we can see what happens if (\ref{psi1k}) is taken to be the bosonic sector of a SUSY partner, i.e. if $\psi_{1,k} = \psi^{+}_{1,k}$.
Then (\ref{partners}) gives its fermionic partner $\psi^{-}_{1,k}$ by applying $Q_1$, viz.
\begin{align}
\begin{split}
\psi^{-}_{1,k} &= \frac{1}{k} Q_1 \psi^{+}_{1,k}
\\
	&= \frac{1}{k}  \left[ \frac{d}{dx} - \frac{1 + 1/2}{x} \right] \psi_{1,k}(x) 
\\
	&= A_{1,k} \sqrt{x} \left[ - J_2(kx) + \vare \pi \left( \tfrac{1}{2} k L \right)^2 N_2(kx) \right]	
\end{split}
\end{align}
This is the general solution further inside the strongly-repulsive range, with $\n = 2$, $\a = \frac{15}{4}$. Here there is no ambiguity, the function $\sqrt{x} N_2(kx)$ also diverges at $x = 0$, and we must set $L = 0$ to regularize both $\psi^{\pm}_{1,k}$ at the same time.

\section{Examples and generalizations}	\label{sec:Apli}

Any $V(x)$ which has the \emph{asymptotic} form of a inverse square potential near $x = 0$ must be subject to the same renormalization process described in Sect.\ref{SectRenor}, irrespective of its form at large $x$.
As a consequence, the anomalous scale $L$ may appear in these theories.
As we give the three examples below, we would like to call attention to a generalization of the method described in Sect.\ref{sec:SUSY}.
If $V(x)$ is supersymmetric, i.e. if it has the form (\ref{SUSYV1}),  and if SUSY is spontaneously broken, we can use the same procedure as above to fix the anomalous scale, even if (unlike the inverse square potential) $V(x)$ is not shape invariant under SUSY.

\subsection{The radial motion of a free particle}

The most basic appearance of the inverse square potential is in the radial Schr\"odinger equation for a free particle,
\be
\frac{2m}{\hbar^2}V(x) = \frac{\ell (\ell + 1)}{x^2}, \quad \ell \in\mathbb{N}, \quad x>0.	\label{livre}
\ee
For every $\ell \geq 1$, the number $\a_\ell = \ell ( \ell +1)$ lies in the strongly repulsive regime (\ref{StrgRepRang}), but the $s$-wave, with $\ell = 0$, lies in the weak-medium range, $\a_0 = 0$.
Of course, there is no physical reason for the existence of a bound state associated with a scale $L$ --- this is simply a free particle. In fact, there is no renormalization needed at all since there are no interactions; the singularity at $x = 0$ is just a problem of the spherical coordinate system.

Nevertheless, this gives an interesting illustration of how the use of SUSY solves paradoxes introduced by a (here forceful) renormalization. The superpotential associated with (\ref{livre}) is 
$(\sqrt{2m} / \hbar) W_\ell (x) = - (\ell + 1) / x$ and the wave-function for $\ell = 0$ with $B_{0,k} = 0$ is 
\be
\psi_{0,k}(x) = A_{0,k} \sqrt x J_{1/2}(kx).
\ee
Following the steps of Sect.\ref{sec:SUSY}, this function generates the solution for every $\ell$ by applying multiple operators $Q_\ell$ constructed from the chain of partner models with degenerate spectra:
\be
 \psi_{\ell,k} (x) = k^{-\ell} Q_\ell Q_{\ell-1} \cdots Q_1 \psi_{0,k} (x) \sim \sqrt{x} J_{\ell + 1/2}(k x) .
\ee
In this particular case, the action of the $Q$ operators is equivalent to a recurrence relation between spherical Bessel functions, see e.g. \cite{Landau_3}.

\subsection{The generalized Calogero-Moser-Sutherland potential}

A non-trivial example is the (shifted) Calogero-Moser-Sutherland potential\cite{gibbons1984generalisation,wojciechowski1985integrable}%
\footnote{%
This is an hyperbolic generalization of the Calogero-Moser-Sutherland trigonometric potential \cite{sutherland1971exact}, which is the subject of the next example. Cf. \cite{Olshanetsky:1983wh}.}
\be
\frac{2m}{\hbar^2}V_+(x) = \frac{\nu^2-1/4}{\sinh^2(x/a)}+(\nu+1/2)^2, \quad  x>0,		\label{PoschlTeller}
\ee
with $0<\nu<1$ (we exclude $\nu=0$ for simplicity). 
The solution of (\ref{Schr}) is given by hypergeometric functions,
\begin{align}
\begin{split}
&\psi_{\n,k}^{+}(x) = a^{\frac{1}{2}} \left[ \sinh \left( \tfrac{1}{a} x \right) \right]^{\n+\frac{1}{2}}  \left[\cosh\left( \tfrac{1}{a} x\right)\right]^{\frac{3}{2}} 
	 \Big[ \tfrac{ (a/2)^\n }{\nu\Gamma(\nu)} k^\n A_{\n,k} \, F \left[ \omega, \bar\omega ; 1+\n ; -\sinh^2\left( \tfrac{1}{a}x \right) \right] 
\\
&\quad\qquad - \left[\sinh\left(\tfrac{1}{a} x\right)\right]^{-2\nu} \tfrac{\Gamma(\nu)}{\pi (a/2)^\n } k^{-\n} B_{\n,k} 	F \left[ \omega - \n, \bar \omega - \nu, 1-\nu ; -\sinh^2\left(\tfrac{1}{a}x\right)\right] \Big]
\end{split}\label{psi_sinh}
\end{align}
where $\omega \equiv 1-\frac{\nu}{2}-\frac{i}{2}\sqrt{2mEa^2/\hbar^2-(\nu+1/2)^2}$. 
The integration constants where chosen such that the asymptotic forms of (\ref{psi_sinh}) and of (\ref{sol}) coincide near the singularity $x = 0$.

Here renormalization is indeed necessary, and results in the ratio $B_{\n,k} / A_{\n,k}$ being given by Eq.(\ref{BAneq0}).
The model is described by the superpotential
\be
\frac{\sqrt{2m}}{\hbar} W(x) = -\frac{1}{a}\left(\nu+ \tfrac{1}{2} \right)\coth\left(x/a\right) .	\label{W_sinh}
\ee
The solution of the zero-mode,
$\psi^{+}_{\n,0}(x)\sim [\sinh(x/a)]^{\nu+1/2}$,
obtained from solving $Q\psi^{+}_{\n,0}(x)=0$,
is not normalizable. Therefore the spectrum of (\ref{PoschlTeller}) is the same as that of its superpartner
\be
\begin{split}
\frac{2m}{\hbar^2}V_- (x) &=\left(\frac{\sqrt{2m}}{\hbar}W(x)\right)^2+\frac{\sqrt{2m}}{\hbar}\frac{dW(x)}{dx}
	=\frac{(\nu+1)^2-1/4}{\sinh^2(x/a)}+(\nu+1/2)^2.
\end{split}
\ee
Near the origin, we find the inverse square potential with a strongly repulsive coupling,
 \[
 (2m/\hbar^2)V_- (x)\approx [(\nu+1)^2- \tfrac{1}{4} ]/x^2.
 \]
 There is no renormalization in this model, and since
 $\psi_{\n,k}^{-}=(1/k)Q\psi^{+}_{\n,k} (x)$, we must fix $L=0$, i.e. $B_{\n,k}=0$ in (\ref{psi_sinh}).

\subsection{The Calogero-Moser-Sutherland potential}

With this last example we will show that, in potentials which are asymptotically inverse square, SUSY restricts the energy spectrum more than the normalizability condition alone. 
It fixes $L = 0$ even when there are square-integrable solutions at the IR fixed point where $L \to \infty$.

The Calogero-Moser-Sutherland potential \cite{sutherland1971exact} and its superpotential are
\begin{align}
\frac{\sqrt{2m}}{\hbar}W(x) &= -\frac{\pi}{a}\frac{(\nu+1/2)}{\sin(\pi x/a)}, \quad 0 < x < a 	\label{W_sin},
\\
\frac{2m}{\hbar^2}V_+(x) &= \frac{\pi^2}{a^2}\frac{(\nu+1/2)}{\sin^2(\pi x/a)} \left[ \nu + \tfrac{1}{2}-\cos\left(\tfrac{\pi}{a} x \right) \right] , 		\label{V_sin}
\end{align}
with $0 < \n < 1$. 
The potential is an infinite well, with inverse square behavior at the boundaries. 
Near $x = 0$, it goes as
$(2m/\hbar^2)V_+ (x)\approx (\nu^2-1/4)/x^2$
which has a medium-weak coupling. 
Near $x = a$, it goes as 
 \[
 (2m/\hbar^2)V_+ (x)\approx \frac{(\nu+1)^2- \frac{1}{4}}{(a-x)^2},
 \]
 with a strongly repulsive coupling.
 
The general solution for the stationary wave function is ($k=\sqrt{2mE/\hbar^2}$) 
\begin{align}
\begin{split}
&\psi_{\n,k}^{+}(x)= \left[ \sin\left(\tfrac{\pi}{2a} x\right) \right]^{\n + \frac{1}{2}}  \left[ \cos \left(\tfrac{\pi}{2a} x \right) \right]^{\nu+ \frac{3}{2} } 
\\
& \times \Bigg[ C^\UV_{\n ,k} \, F \left[ 1+ \nu -\tfrac{ka}{\pi} , 1+ \nu + \tfrac{ka}{\pi} ; 1 + \nu ; \sin^2\left(\tfrac{\pi x}{2a} \right) \right]
\\
& \qquad + C^\IR_{\n,k} \left[ \sin\left(\tfrac{\pi x}{2a}\right) \right]^{-2\nu}   F \left[ 1 -\tfrac{ka}{\pi} , 1 + \tfrac{ka}{\pi} ; 1-\nu ; \sin^2\left(\tfrac{\pi x}{2a}\right) \right] \Bigg] .
\end{split}
\end{align}
Near  $x=0$, as expected, $\psi_{\n,k}^{+}(x)\approx f_\UV (x) + f_\IR (x)$, 
where
 $f_\UV (x) \propto C^\UV_{\n ,k} \, x^{\nu+\frac{1}{2}}$ 
and 
$f_\IR(x) \propto C^\IR_{\n ,k} \, x^{-\nu+ \frac{1}{2}}$.
Both solutions as square-integrable at the origin, and the renormalization procedure fixes the ratio
$
C^\IR_{\n ,k} /  C^\UV_{\n ,k} \propto L^{2\nu}. 
$
 On the other hand, \emph{neither} of the solutions is square-integrable in $x = a$, unless one of the hypergeometrics is a polynomial --- which gives a discrete condition on $k$ --- and the other solution is discarded by fixing the respective $C = 0$, i.e. by choosing one of the fixed points. 
 The discrete spectra on the UV point ($L = 0$) and on the IR point ($L = \infty$) are, 
 \begin{align}
&& k^\UV_{\n, n} &= \frac{\pi}{a}(n+1+\nu), &&\quad C^\IR_{\n ,k} = 0, &&	\label{k_UV}
\\
&& k^\IR_{\n, n} &= \frac{\pi}{a}(n+1), &&\quad C^\UV_{\n ,k} = 0, &&	\label{k_IR}
\end{align}
where $n \in\mathbb{N}$.

We therefore have two classes of normalized, renormalized eigenfunctions, each class with a different spectrum. 
Renormalization (or the self-adjoint extension) around $x = 0$ tells us that the two spectra cannot coincide, since they correspond to two different fixed points, but it does not give any clear criterion for the preference of one over the other. It is SUSY that chooses the spectrum unambiguously. 

Indeed, the partner potential of (\ref{V_sin}) is
\be
\frac{2m}{\hbar^2}V_- (x) = \frac{\pi^2}{a^2}\frac{(\nu+1/2)}{\sin^2(\pi x/a)} \left[ \nu+ \tfrac{1}{2} +\cos \left( \tfrac{\pi}{a} x \right) \right].		\label{V_sin_par}
\ee
Near the origin, $V_-(x) \sim [(\nu+1)^2-1/4]/x^2$,
with a strongly repulsive coupling, while near $x = a$ it goes as $(\nu^2-1/4)/(a-x)^2$. 
In fact, the partner potentials (\ref{V_sin}) and (\ref{V_sin_par}) are the same, they are simply reflected about  $x=a/2$. 
By itself, solving the Schr\"odinger equation for $V_-(x)$ leads to the same ambiguity for the spectrum, but the ambiguity is resolved after we relate the partner wave functions. 
For the UV case, we have
\be
\psi^{- \UV}_{\n, n}(x) = \frac{1}{k_n} Q \, \psi^{+ \UV}_{\n, n}(x) \sim C^\UV_{\n ,n} x^{\nu+ \frac{3}{2}} [1 + \mathcal{O}(x^2) ]. 
\ee
It is possible to show that  $\psi^{- \UV}_{\n , n}$ is normalized if $\psi^{+\UV}_{\n , n}$ is normalized, and both generate the same spectrum (\ref{k_UV}). 
Meanwhile, the fermionic partner of IR solution
\be
\psi^{- \IR}_{\n,n} \propto Q \, \psi^{+ \IR}_{\n ,n}(x) \sim C^\IR_{\n ,n} \, x^{-\nu - \frac{1}{2}} [ 1+\mathcal{O}(x^2)] ,
\ee
is clearly not square-integrable at the origin. 
Therefore the IR spectrum is not consistent with SUSY, which selects the UV fixed point.

\section{Discussion}\label{sec:Con}


Our  main result is the proof  that there is a SUSY in the parameter space of inverse square potentials with coupling $\a$. It relates the weak-medium ($- \frac{1}{4} \leq \a < \frac{3}{4}$) and the strongly-repulsive ($\a \geq \frac{3}{4}$) ranges of the potential. As a consequence of this relation,  SUSY kills the anomalous scale that appears in the quantization of the weak-medium range, by setting $L = 0$ and forbidding the renormalized theory to leave the UV fixed point of its RG flow.
Meanwhile, in the strongly-attractive range, there is no SUSY description, since the would-be partner-potentials of (\ref{IinvSqPot}) for $\a < -\frac{1}{4}$ become complex functions and the Hamiltonians are non-Hermitian. 

Hence \emph{dynamical SUSY is a property of the the potential (\ref{IinvSqPot}) only before and at the BKT-like phase transition; it disappears completely at the discrete-scaling phase}. 
We consider the existence of a dynamical SUSY, by itself, to be a noteworthy fact about the renormalized Hamiltonian (\ref{IinvSqPot}).

The inverse square potential with a weak-medium coupling appears in many different contexts, and our argument for fixing $L = 0$ results in rather non-trivial consequences.
The first example is the radial scattering of a charged, non-relativistic, spinless particle by a thin solenoid (the Aharonov-Bohm effect). One of the present authors has shown that in this scenario up to two phase shifts have to be renormalized \cite{CamaradaSilva:2018leo}, introducing up to two anomalous quantum scales. The results of the present paper, however, strongly suggest that these scales should be set to zero, thus recovering the usual formula for the cross-section after supersymmetry is imposed.

The Schr\"odinger equation (\ref{Schr}) also describes fluctuations of asymptotically Anti-deSitter (AdS) domain walls. In the gauge/gravity correspondence, the spectrum of (\ref{IinvSqPot}) is related to the mass spectrum of particles living the AdS boundary and controlled by the bulk geometry which can be singular \cite{Gursoy:2007er,Kiritsis:2006ua,Kiritsis:2016kog}.
In this context, the ambiguity of the renormalized solutions in the weak-medium coupling range, here codified in the scale $L$, are related to the necessity of assigning holographic boundary conditions at a singularity (not at the AdS boundary), which is an unphysical situation.
In a recent paper \cite{Lima:2019bsi}, three of the present authors have shown that the SUSY transformation for the fluctuations corresponds to a symmetry of the bulk $d+1$-dimensional domain wall relating large and small scales. With the appropriate translation, the same arguments presented here can be used to fix $L = 0$, then fix the ambiguities in the boundary conditions discussed in \cite{Gursoy:2007er,Kiritsis:2006ua,Kiritsis:2016kog}.

\appendix

\section{Renormalized solutions for $\a \leq - 1/4$}	\label{AppA}

In the strongly repulsive regime, the Bessel index becomes imaginary, $\nu= i \tilde \nu$ with $\a = - ( \tilde \n + \frac{1}{4})$.
The renormalized zero-energy solution is 
\be
\psi_{\tilde \nu;0}(x) = C_{\tilde \n}(0) \sqrt{x}\sin\big[ \tilde \nu\ln(x/\tilde L)+\delta \big]\label{Eeq0_2},
\ee
with a length scale $\tilde L$  and a dimensionless integration constant $\delta$ which cannot be fixed by a boundary condition. 
By the same steps as before, one finds the running coupling \cite{Braaten:2004pg}
\be
\gamma(R) 
		= \tilde\nu \cot \left[ \tilde\nu \log (R/\tilde L)+\delta \right] .	\label{gamma_2}
\ee

The shallow bound states referred to in the main text can be found by looking at solutions for $E < 0$ which are regular at $x = \infty$; we find
\begin{align*}
\begin{split}
&\psi_{\tilde \nu}(x) \propto K_{i \tilde \nu}(\kappa x) 
	\sim \sin\left[ \tilde \nu \log (\kappa x/2)+\theta_{\tilde \nu} \right] , 
	\\
&\kappa=\sqrt{-2mE/\hbar^2}, \quad \theta_{\tilde \nu} = \frac{1}{2i} \log \left[\Gamma(1- i\tilde \nu) / \Gamma(1+i \tilde\nu)\right]
\end{split}
\end{align*}
where we used $\kappa x\ll1$. Continuity then results in
\be
\kappa_n = \kappa_0 e^{-\frac{\pi}{\tilde \nu}n}, 
\quad  
\kappa_0 \equiv (2 / \tilde L)e^{(\delta-\theta_{\tilde \nu})/\tilde \nu}, \qquad n \in \mathbb{Z}.
\ee
This is the discrete energy spectrum $E_n \sim - \kappa_n^2$.


\bibliographystyle{utphys}

\bibliography{SUSYshieldsScalingRef} 

\end{document}